\documentclass[runningheads]{llncs}
\usepackage{graphicx}

\usepackage{times}
\usepackage{epsfig}
\usepackage{amsmath}
\usepackage{amssymb}
\usepackage{subfigure}
\usepackage{cite}

\begin{document}
\title{Synthetic Generation of Three-Dimensional Cancer Cell Models from Histopathological Images}
%\title{Transformers for Synthetic Generation of Three-Dimensional Cancer Cell Models from Histopathological Images}
%\title{Deep Topological Transformer for Context Aware Generation of Cancer Cell Clusters}
%Deep topological transformer for context aware generation of cell clusters
\titlerunning{Synthetic Generation of 3D Cancer Cell Models from Histopathological Images}
\author{Yoav Alon\inst{1} \and
Xiang Yu\inst{1} \and
Huiyu Zhou\inst{1}
}
\authorrunning{Y. Alon et al.}
\institute{School of Informatics, University of Leicester, UK 
\\
\email{\{ya88,xy144,hz143\}@leicester.ac.uk}}
\maketitle
\begin{abstract}
Synthetic generation of three-dimensional cell models from histopathological images aims to enhance understanding of cell mutation, and progression of cancer, necessary for clinical assessment and optimal treatment. Classical reconstruction algorithms based on image registration of consecutive slides of stained tissues are prone to errors and often not suitable for the training of three-dimensional segmentation algorithms. We propose a novel framework to generate synthetic three-dimensional histological models based on a generator-discriminator pattern optimizing constrained features that construct a 3D model via a Blender interface exploiting smooth shape continuity typical for biological specimens. To capture the spatial context of entire cell clusters we deploy a novel deep topology transformer that implements and attention mechanism on cell group images to extract features for the frozen feature decoder. The proposed algorithms achieves high quantitative and qualitative synthesis evident in comparative evaluation metrics such as a low Frechet-Inception scores.

\end{abstract}

\section{Introduction}
The clinical assessment process for breast cancer involves initial imaging in form of mammography, computer tomography, or ultrasound, followed by a medical evaluation determining the need to conduct a biopsy. Final confirmation for the presence of cancer can only be given based on histopathological analysis of cell tissue \cite{BreastHistology}, as the pathological cell mutation is considered the most determining criteria of cancer \cite{GeneralDeepLearning, RecurrentAttention}.
For the histopathological examination, surgically extracted tissue undergoes several steps of staining and fixation, and the resulting histopathological images can be digitized as whole slide images to form a pyramid of different resolution scans. The significant advantage of spatial visualization of cell topology for understanding physiological patterns and thereby improving accurate diagnosis has been demonstrated in previous works \cite{3DReconstruction, 3DReconstructionMultiple, MicebrainReconstruction, Reconstruction3DHistology}. Three-dimensional histopathological data is usually based on the reconstruction of consecutive slides of stained tissue. Together with common challenges of reconstruction such as artifacts and cell deformation, three-dimensional datasets are difficult to obtain and often not publicly available. The complex annotation process of three-dimensional data poses an additional challenge where reconstruction often leads to inconsistency between slide annotations.

Cutting-edge cancer segmentation algorithms require a large set of robust three-dimensional data to optimize cancer segmentation ability for various types of tissue. Available datasets usually focus on a specific type of tissue and generated models cannot be generalized for other kinds of cancer. The reconstruction specific artifacts further harm the segmentation algorithms ability to identify relevant features. We, therefore, strive to generate synthetic datasets that are suitable to train segmentation models without losing spatial context information and cell-specific features. The essence of a histopathological image is that the two-dimensional surface encodes all the information necessary to reproduce a three-dimensional image. We propose a novel algorithm that generates three-dimensional cell models based on 2D histopathological images and then uses the cross-sections to distinguish between the original and the generated cell images. We aim to achieve accurate generation of cell cores and membranes and to describe an accurate representation of synthetic cell topology that enables the training of algorithms for the segmentation of cancer areas and the classification of cell mutation stages. Additionally, the synthesis of cell clusters requires a neural network to learn the topological context of different cell types. The clinical knowledge as result of learned generalization may be valuable in understanding the development of cancer. The generation of synthetic 3D data is not only an acceptable compensation for missing real-life data but also an accurate augmentation tool. 

Our approach encompasses the following contributions:
\begin{itemize}
  \item A synthetic three-dimensional cell generation framework based on a generative adversarial architecture. 
  \item A novel deep topology transformer that implements an attention mechanisms to extract spatial context using a encoder, which allows for the generation of entire cell clusters taking into account their typical topological context.
  %\item The generation of synthetic cell clusters using a style transfer network for synthetic background generation. 
\end{itemize}

\section{Related Work}
A comprehensive survey of recent methods in 3D histology reconstruction was presented by Pichat \cite{3DReconstruction}, which portrays the issue of invasive histology where cell group topology is broken due to the cutting of tissue slides in several slides. A reconstruction that enables the spatial visualization of tissue requires {\it a prior} understanding of normal tissue shapes. The reconstruction from coherent macroscopic features to microscopic properties with spatial correspondence can be achieved through image registration. 

Comparative analysis of tissue reconstruction algorithms for 3D histology, reported in \cite{Reconstruction3DHistology}, discusses the reconstruction based on serial tissue selections. Their approach forms a benchmarking framework for the evaluation of the existing reconstruction algorithms, where reconstruction performance mainly depends on the capability of system compensation for local tissue deformation. The approach of \cite{3DReconstructionMultiple} focuses on 3D tissue reconstruction using different stains in histology images. A quantitative visualization tool in their system enables spatial alignment of structural and functional elements. Growth patterns and spatial arrangement of cells can improve the understanding on a physiological and pathological level and enable better treatment. A lower overall error in reconstruction has been achieved using their proposed alignment strategy. 
A prominent application for 3D reconstruction is given in \cite{MicebrainReconstruction}, which reconstructs a high resolution 3D anatomical reference atlas of the mouse brain from histological sections. A combination of high-frequency slice-to-slice histology and low-frequency histology-to-MRI registration produces an accurate reconstruction in terms of shape and alignment of local features. 
3D generative adversarial modeling has been proposed in \cite{3dSynthesis} to learn a probabilistic latent space of object shapes. Using recently developed volumetric convolutional networks and modern generative adversarial networks, 3D objects are created from a representation space. The generator network makes use of volumetric fully convolutional layers. The resulting discriminator can additionally serve as a highly accurate classifier for 3D objects. Two 3D object generation algorithms 3D-GAN and 3D-VAE-GAN were proposed and analytically compared. 
Similarly, the generation of three-dimensional structures has been achieved by 3D shape induction from 2D views of multiple objects in \cite{3DShapeInduction}. Protective generative adversarial networks (PrGANs) were developed to match the projections of 3D objects against the 2D input images. Novel views can be created from any input image in an unsupervised manner. Accordingly, the discriminator uses common 2D convolutional layers, where the generator inspired by 3D-GAN \cite{3dSynthesis} uses volumetric fully convolutional layers whilst the generator only creates a 3D voxel representation. 

The original Generative Adversarial Networks (GANs) are based on an adversarial process of training two models, a generator and a discriminator, that estimates the probability for a sample to be an original or a generated image \cite{GAN}. The generator has to achieve a high probability for the discriminator to make a mistake in the estimation. Since their establishment, GANs have been used in a wide variety of applications such as computer vision \cite{GansComputerVision} and many others. A survey of recent GAN variants and applications can be found at \cite{GANSurvey}.
A variant of GANs that avoids the common mode collapse problem, i.e. creation of very similar images that satisfy the discriminator, is referred to as Wasserstein GAN \cite{WassersteinGan} which is defined on the hypersphere manifold. 

%To fully take advantage of the projection of complete cell clusters, we may use a variant of CycleGan \cite{CycleGan, CycleGanGit, ImageToImage}, an unpaired image-to-image translation based on two simultaneously trained generators and two discriminators. In these approaches, cycle consistency suggests that similarity to an original image should be achieved after we pass it through both generators. 

%More insight into cell mutation and essentials of cell biology can be found at \cite{CellMutation, CellBiology}. In brief, cancer is caused by a genetic mutation that accelerates cell division rates or modifying cell control systems as cycle arrest or cell death. Signals that regularly control cellular growth and death are not adequately responded to. With ongoing cell growth, the deviation from regular cells becomes more drastic. A group of growing cells forms a tumor that might invade neighboring tissues. 
The application of deep learning including detection and segmentation of cancerous cells in histology has been discussed in \cite{MLHisto, MLHisto2, MLHisto3, MLHistoImageAnalysis}. 

\begin{figure}[]
  \centering
  \subfigure{\includegraphics[height=6.05cm]{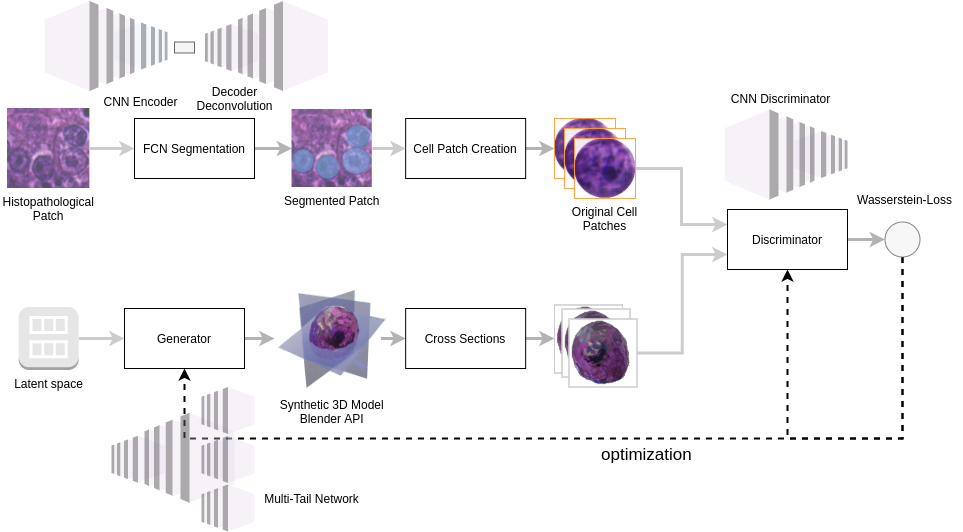}}\quad
  \caption{The proposed workflow includes cell segmentation for cell patches using a fully connected network and the extraction of resulting blobs. On the generator site, a latent space vector is converted to feature vectors that are passed as parameters to the blender API with explicit constraints for generating a synthetic 3d model of cells. Projections of the model are then used to compare between the original and generated cell images, and then optimization is applied to improve the generation of the synthetic three-dimensional cell model. The background of both, the original and generated cells, is transparent to avoid  distinguishing patches based on background indications by the discriminator.}
  \label{architecture}
\end{figure}

\section{Proposed Method}
%Our approach incorporates three components. First, the synthetic three-dimensional cell generation forms the basis of all further work, then the training of an embedding space reveals cell mutation stages, and finally the generation of entire synthetic cell clusters. 
%\subsection{Cell synthesis}
%The foundation of the synthetic generation of three-dimensional cancer cell models is the original set of two-dimensional histopathological images that contain a large number of segmentable cells. The differentiation between cancerous and normal cells is based on an area-wise annotated ground truth prepared under the supervision of expert pathologists \cite{Cameleyon17}. 
\subsection*{Synthetic cell model by feature generation}
For the synthetic generation of a single three-dimensional cell model we construct an architecture (See Figure \ref{architecture}) based on the generative adversarial network pattern that generates features of a three-dimensional model and uses it's projections to train a discriminator to distinct original and generated images. Background-free cell patches are extracted by a simple fully connected segmentation network from histopathological images and serve as original samples. 
%Detailed discussion on cell segmentation algorithms is out of the scope of this manuscript, but a solid understanding can be obtained by referring to the approaches of \cite{NeuralCellDetect, CellDetection, SegmentationCellNet, TBNCDataset}. Nevertheless, high-quality detection is required to obtain background free centered cells for the discriminator not to be deceived to distinct original patches based on the remains of background pixels. Our implementation extends a fully fully convolutional network based on the models proposed in \cite{FCNSegmentation} and \cite{NeuralCellDetect}.
The proposed architecture (See Figure \ref{architecture}) resembles the regular GAN pattern, but the generator network architecture incorporates multiple tails that are fed as feature vectors to the blender bpy API script. The extracted features describe a deformation tensor of cells and membranes, a surface tensor containing distance and strength information of a blender modifier and surface pixels, relative positions of cells and membranes, and color information. The features are constrained to avoid non-continuous singularities and to create realistic cell mutations. Thus, compared to volumetric convolutional \cite{3dSynthesis} or a voxel generation approach \cite{3DShapeInduction}, constraining the cell form by feature vectors will ensure the generation of biologically feasible models. A custom blender API script processes these features as input to construct a three-dimensional model. Two-dimensional projections are rendered, and multiple cross-sections at various angles $\theta$ and $\phi$ around a center of gravity result in a batch of generated images. The discriminator will be trained to distinguish original and generated images, and at the same time, the generator is optimized to deceive the discriminator. 
To avoid mode co

To avoid mode collapse and achieve stable performance where the generator still optimizes well even when the discriminator has gained a high ability to distinguish between the original and generated samples we extend the Wasserstein loss \cite{WassersteinGan} : 
\begin{equation}
    F_{W_{GAN}} = \nabla_{\theta_d} \frac{1}{m} \sum_{i=1}^m [f(x^{(i)} - f(G(z^{(i)}))] , F_{W_{g}} = \nabla_{\theta_g} \frac{1}{m} \sum_{i=1}^m f(G(z(^{(i)}))
\end{equation}
The distance is a minimum cost of transporting mass in converting a data distribution to another.
\begin{equation}
    W(P_r, P_g) = \inf_{\gamma \in \Pi(P_r, P_g)} \mathop{\mathbb{E}}_{(x,y)\sim \gamma} [||x-y||]
\end{equation}
%For the Wasserstein-GAN, the cost functions are: % is defined by the discriminator's cost: 
%\begin{equation}
%    F_{WGAN} = \nabla_{\theta_d} \frac{1}{m} \sum_{i=1}^m [f(x^{(i)} - f(G(z^{(i)}))] , F_{W_{g}} = \nabla_{\theta_g} \frac{1}{m} \sum_{i=1}^m f(G(z(^{(i)}))
%\end{equation}
%Its generator's cost is: 
%\begin{equation}
%    F_{W_{g}} = \nabla_{\theta_g} \frac{1}{m} \sum_{i=1}^m f(G(z(^{(i)}))
%\end{equation}
The discriminator does not use an output sigmoid function, and the cost function remains different. Additionally, the use of the the above loss function prevents vanishing gradients and is itself able to establish a measure of distance in the space of probability distributions. 
%\begin{figure*}[]
%  \centering
%  \subfigure{\includegraphics[scale=0.23]{images/CycleFramework.png}}\quad
%  \caption{Proposed architecture using unpaired image-to-image translation for style transfer creating synthesized cell-cluster images based on a three-dimensional model projection. We propose to convert a model projection via style transfer to a realistic image, containing the projection of three-dimensional cells in the 2D space with the background tissue as a result of context-aware style-transfer by the CycleGan. During training, an adversarial loss serves to match generated images to the target domains data distribution and a cycle consistency loss prevents contradiction.}
%  \label{CycleGan}
%\end{figure*}
%\subsection{Cell cluster synthesis}
\begin{figure}[]
  \centering
  \subfigure{\includegraphics[height=4.4cm]{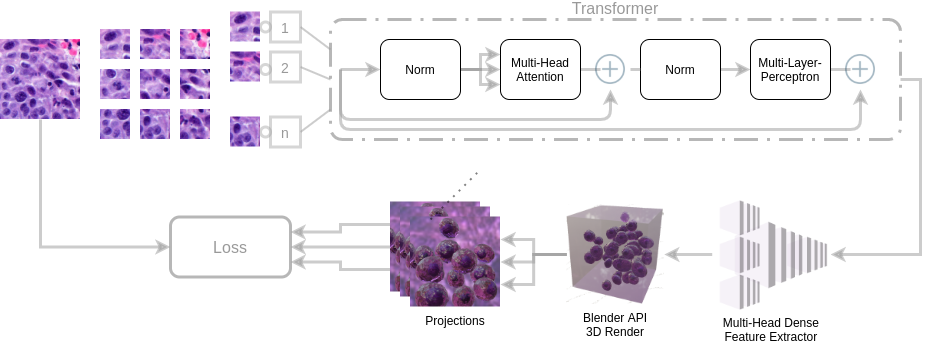}}\quad
  \caption{Transformer for the extraction of topological information using a attention mechanism. The transformer encoder is extended by a (frozen) multi-head decoder that constructs features for the Blender API generation of three-dimensional clusters. Projections are then extracted and the similarity between projections and original is used to optimize the transformer weights similar to auto-encoder loss for the most similar projections. The encoder consists of alternating layers of multi-head self attention and multi-layer perceptrons with applied layer-norm and residual connections. }
  \label{transformer}
\end{figure}

\subsection*{Deep topological transformer for context aware generation of cell clusters}
The synthesis of entire cell clusters requires neural network based understanding of the topological context of cells that vary from different tissue types. Based on adapted transformers for image classification \cite{transformers, ImageTransformer} we device a innovative transformer designed for spatial attention that encodes the probabilistic configurations of cell clusters and provides feature vectors f that are feed into our previous cell generation architecture. 

%method
Original histopathological images in magnification 40 are split into smaller patches that contain several cells. The patches are created from this image and the linear projection of flattened patches is fed into a transformer encoder. To retain positional information we must add position embeddings to patch embeddings. The transformer is then build using alternating layers of multi-headed self attention (MSA) (See Figure \ref{transformer}). As output of the transformer encoder we device a multi-layer perceptron that returns features needed for the Blender API.  The encoder consists of alternating layers of multi-head self attention and multi-layer perceptrons with applied layer-norm and residual connections. 
Using Multi-Head attention queries, keys and values are linearly projected h times with different projections. The attention function is executed in parallel where we concatenate the output values and project them again. Information from different positions is attended using Multi-head attention similar to \cite{ImageTransformer}.

\begin{equation}
M(Q,K,V) = W^O [h_1, h_2, h_n]_c
\end{equation}
with 
\begin{equation}
    h_i = A(QW_i^Q, KW_i^K, VW_i^V)
\end{equation}
where M is the Multi-Head function with inputs Q,K,V and A is the attention function. $W_i^Q$ are projections matrices for their respective inputs.
The reconstruction error ${\cal L_T}\left( {x,\hat x} \right)$  for original $x$ and synthetic projection $\hat x$ now must be determined based on a batch of projections $\langle \hat x(\theta, \phi) \rangle$. The sum of losses for all possible pairs $\langle \hat x(\theta_i, \phi_j) \rangle$ would not be effective since it would not consider the relevant angles. But we should obtain the sum of n minimum reconstruction errors: 

\begin{equation}
{\cal L_T}\left( {x,\hat x} \right) = \sum_{i,j} 
{\cal L}\left( {x,\hat x} \right) \textrm{ with } \min_n  \langle \hat x(\theta_i, \phi_j) \rangle
\end{equation}
for $n$ smallest errors with $i,j$. Then we add a loss term which penalizes large derivatives of our hidden layer activations and get:
%\begin{equation}
%    \min_n  \langle \hat x(\theta_i, \phi_j) \rangle \textrm{ with}
%\end{equation}
\begin{equation}
    {\cal L_T}\left( {x,\hat x} \right) = {\cal L}\left( {x,\hat x} \right) + \lambda {\sum\limits_i {\left\lVert {{\nabla _ x}a_i^{\left( h \right)}\left( x \right)} \right\rVert} ^2}
\end{equation}

\section{Experiments}
Prototype development and initial experiments were performed locally using an Nvidia Tesla 10 8GB, and further experiments were performed using the High-Performance computing (HPC) Cluster of our organization with advanced GPUs. The implementation was realized using PyTorch 1.6 and Python 3.6 together with the Blender 2.90.1 Python API. The source code of this work will be published on GitHub. The original histopathological whole-slide images used in this work include breast cancer metastases of histological lymph node sections on the lesion- and patient-level and are part of the Cameleyon17 dataset  \cite{Cameleyon17}. The annotated data is based on microscopic hematoxylin and eosin (H\&E) stained samples.

\begin{figure}[]
  \centering
  
  \subfigure[]{\includegraphics[height=2.2cm]{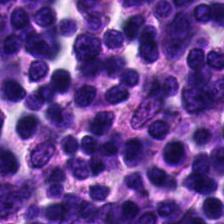}}\quad
  \subfigure[]{\includegraphics[height=2.2cm]{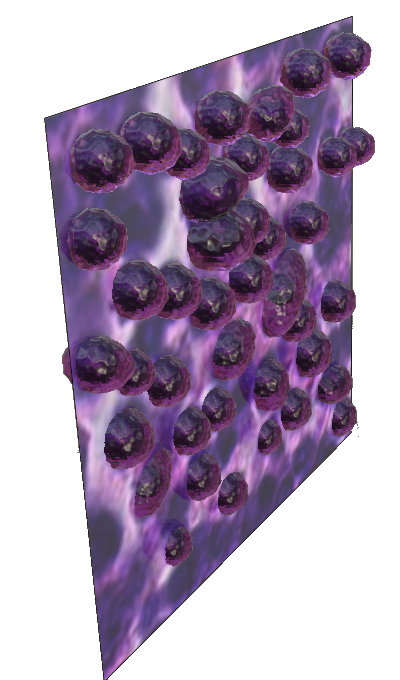}}\quad
  \subfigure[]{\includegraphics[height=2.2cm]{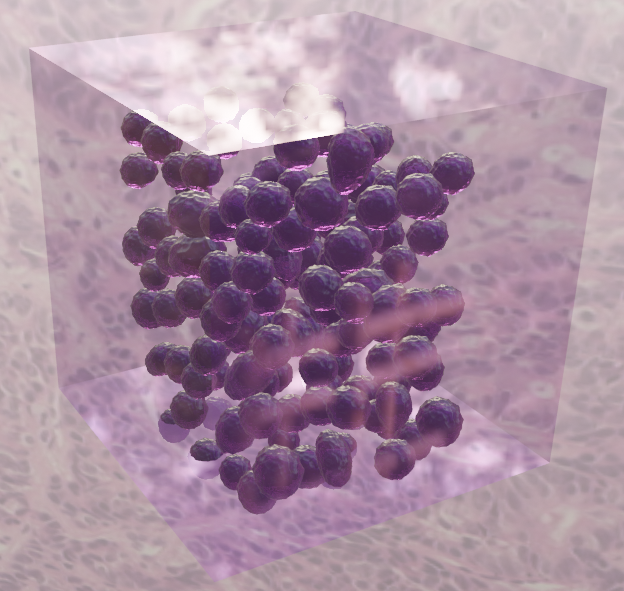}}\quad
  \subfigure[]{\includegraphics[height=2.2cm]{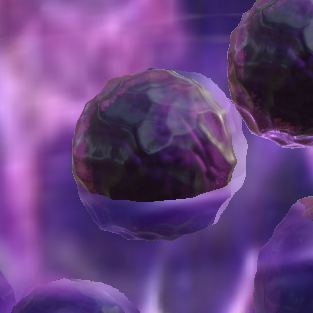}}\quad
  \raisebox{0.6cm}{\rotatebox[origin=t]{90}{Normal}}

  \subfigure[]{\includegraphics[height=2.2cm]{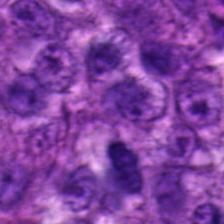}}\quad
  \subfigure[]{\includegraphics[height=2.2cm]{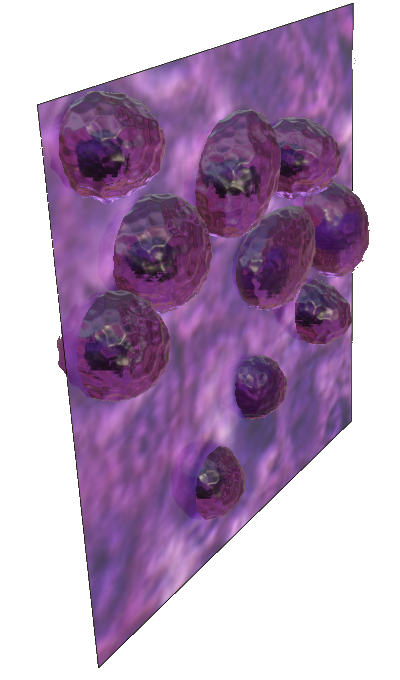}}\quad
  \subfigure[]{\includegraphics[height=2.2cm]{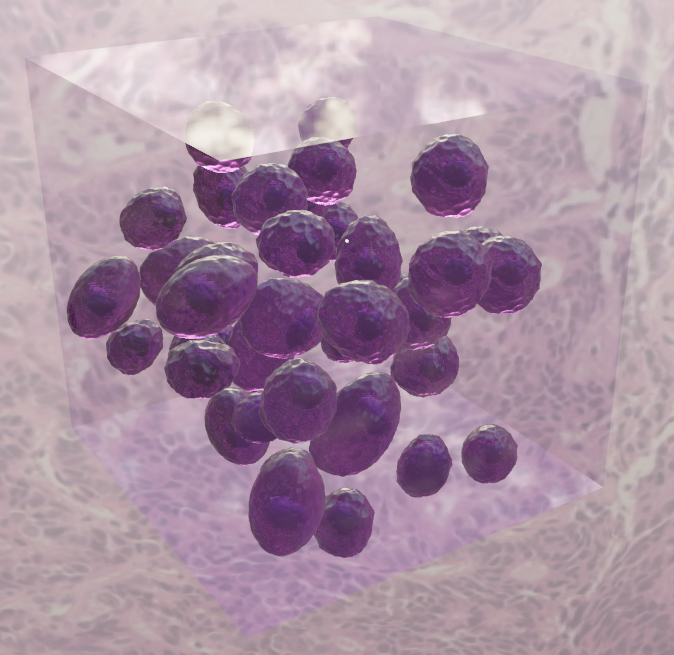}}\quad
  \subfigure[]{\includegraphics[height=2.2cm]{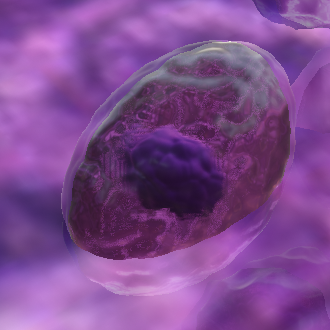}}\quad
  \raisebox{0.6cm}{\rotatebox[origin=t]{90}{Cancer}}

  \caption{Synthetic cell cluster generation for common cell clusters (a)-(d) and malignant cell clusters (e)-(h), generated from respective cell patches. (a) \& (e) Original cell patches, (b) \& (f) synthetic cell generation based on encoded features (c) \& (g) three-dimensional cell clusters, and (d) \& (h) synthetic cells. }
  \label{CellClusterExamples}
\end{figure}

For the segmentation of cancer cell patches in the cell synthesis architecture, we use a pre-trained segmentation network, an extension of a fully convolutional network based on \cite{FCNSegmentation} and \cite{NeuralCellDetect}. Although achieving a high segmentation accuracy a certain number of bad cell patches remains having overlay with other tissue or a high amount of border regions, and effectively, the discriminator can easily recognize original patches by looking for features of faulty patches (e.g. containing background). Overall, the influence of faulty cell patches is negligible. In the cell mode, multiple cross-sections are rendered using the python API and fed back to the discriminator. Initially, the extraction of cross-sections by automated rendering emerged to be very time-consuming. Considering the rendering process as a bottleneck, we adapt rendering properties, such as deactivation of screen-space reflections, etc. Various versions of the generators were examined since the number of tail entities creating feature vectors requires extensive experiments with different network configurations or the number of layers and neurons. 

%In most deep learning algorithms, a loss function is trained until convergence is reached. However, for the generative adversarial network, performance evaluation cannot be easily determined only based on the loss function development. Instead, the relative loss is an indication of the ability to misdirect the discriminator. 
Together with visual evaluation, the quantitative measures of Inception \cite{InceptionPaper} based Inception Score \cite{InceptionScore} and Frechet Inception Distance \cite{FrechetMeasure} can be used to quantitatively evaluate the quality of the generated images according to the created models. The analysis of evaluation measures can be found in \cite{EvaluationMeasures}.
The Frechet inception score is defined as: 
\begin{equation}
\text{FID} = ||\mu_r - \mu_g||^2 + \text{Tr} (\Sigma_r + \Sigma_g - 2 (\Sigma_r \Sigma_g)^{1/2}),
\end{equation}
where $X_r \sim \mathcal{N} (\mu_r, \Sigma_r)$ and $X_g \sim \mathcal{N} (\mu_g, \Sigma_g)$ indicate the activation of an inception3 layer. It indicates the quality of generation. A low score stands for a high similarity between the original and the generated images. Quantitatively, good three-dimensional synthesis refers to low FID scores for the rendered projections, compared to the segmented patches. A lower bound of the FID score is given by the nature of the constraints. The determination of FID scores is based on the generated model of a training procedure that satisfies the subjective visual evaluation criteria.
\begin{table}[]
\centering
\begin{tabular}{ll|l|l|l|l|l|l|l|l|}
\cline{3-10}
                                                                                         &                                                                  & \multicolumn{4}{l|}{\textbf{Single-Cell}}                             & \multicolumn{4}{l|}{\textbf{Cell-Cluster}}                            \\ \hline
\multicolumn{1}{|l|}{\textbf{\begin{tabular}[c]{@{}l@{}}Constrained\\ Features\end{tabular}}} & \textbf{\begin{tabular}[c]{@{}l@{}}Tail\\ Entities\end{tabular}} & \multicolumn{3}{l|}{\textbf{FID}}                   & \textbf{$\sigma$(FID)} & \multicolumn{3}{l|}{\textbf{FID}}                   & \textbf{$\sigma$(FID)} \\ \hline
\multicolumn{1}{|l|}{}                                                                   &                                                                  & \textit{Total} & \textit{Regular} & \textit{Cancer} & \textit{}       & \textit{Total} & \textit{Regular} & \textit{Cancer} &                 \\ \hline
\multicolumn{1}{|l|}{5}                                                                  & 4                                                                & 6.311          & 5.194            & 7.428           & $\pm$ 0.07            & 7.351          & 6.756            & 7.946           & $\pm$ 0.1             \\ \hline
\multicolumn{1}{|l|}{32}                                                                 & 4                                                                & 4.381          & 3.846            & 4.916           & $\pm$ 0.08            & 5.610          & 5.044            & 6.176           & $\pm$ 0.09            \\ \hline
\multicolumn{1}{|l|}{1165}                                                               & 4                                                                & 4.194          & 3.899            & 4.489           & $\pm$ 0.07            & 5.224          & 4.414            & 6.034           & $\pm$ 0.09            \\ \hline
\multicolumn{1}{|l|}{4129}                                                               & 5                                                                & 3.945          & 3.756            & 4.134           & $\pm$ 0.06            & 4.879          & 4.186            & 5.572           & $\pm$ 0.08            \\ \hline
\end{tabular}
\caption{Generative cell cluster synthesis based on segmented cell patches. The Frechet-Inception Score gives an indication of the quality of the generated model. A low score stands for a high similarity between the original and the generated images. \textbf{Cell cluster synthesis} is compared against \textbf{single-cell synthesis} using equivalent constrained features and tail entities. The higher complexity of synthesizing entire cell clusters contributes to the performance decreases in terms of the quantitative FID-Score.}
\label{SingleCellTable}
\end{table}

\begin{table}[]
\centering
\begin{tabular}{l|l|l|l|l|l|l|l|l|l}
\cline{2-9}
                                      & \multicolumn{4}{l|}{\textbf{Single Cell}}                             & \multicolumn{4}{l|}{\textbf{Cell-Cluster}}                            &  \\ \cline{1-9}
\multicolumn{1}{|l|}{\textbf{Method}} & \multicolumn{3}{l|}{\textbf{FID}}                   & \textbf{$\sigma$(FID)} & \multicolumn{3}{l|}{\textbf{FID}}                   & \textbf{$\sigma$(FID)} &  \\ \cline{1-9}
\multicolumn{1}{|l|}{}                & \textit{Total} & \textit{Regular} & \textit{Cancer} &                 & \textit{Total} & \textit{Regular} & \textit{Cancer} &                 &  \\ \cline{1-9}
\multicolumn{1}{|l|}{Volumetric \cite{3dSynthesis}}      & 4.9325         & 4.201            & 5.664           & $\pm$ 0.15            & 6.2435         & 5.645            & 6.842           & $\pm$ 0.15            &  \\ \cline{1-9}
\multicolumn{1}{|l|}{Voxel \cite{3DShapeInduction}}           & 7.271          & 6.894            & 7.648           & $\pm$ 0.36            & 7.0125         & 7.649            & 9.376           & $\pm$ 0.42            &  \\ \cline{1-9}
\multicolumn{1}{|l|}{Encoder-Decoder} & 7.4385         & 6.229            & 8.648           & $\pm$ 0.48            & 9.399          & 8.648            & 10.150          & $\pm$ 0.51            &  \\ \cline{1-9}
\multicolumn{1}{|l|}{Reconstructed \cite{MicebrainReconstruction}}           & 9.1525         & 8.173            & 10.132          & $\pm$ 0.55            & 13.0345        & 12.463           & 13.633          & $\pm$ 0.63            &  \\ \cline{1-9}
\multicolumn{1}{|l|}{ours}            & 3.945          & \textbf{3.756}   & \textbf{4.134}  & $\pm$ 0.06           & 4.879          & \textbf{4.186}   & \textbf{5.572}  & $\pm$ 0.08            &  \\ \cline{1-9}
\end{tabular}
\caption{Comparison with most recent approaches in three-dimensional synthesis applied on cell synthesis for regular and cancer cells.}
\label{SingleCellTable}
\end{table}

%In cell cluster synthesis, the lack of the original background free cluster patches has to be addressed. An attempt to create such patches normally fails because with a higher number of cells, the chance of the segmentation being faulty is larger, and as a result, the discriminator would be able to distinguish the features of the original and the generated images based on the remains of the background fragments. For this reason, we use style transfer based on the standard CycleGAN (see Figure \ref{CycleGan} to create realistic projections from our three-dimensional model. Regular unsegmented cell patches serve as input and a generator similar to the previously described architecture is deployed. 

%The style transfer is used to create a synthetic background for the projection of the three-dimensional cell cluster model. 
%The intensive cluster rendering and networks in need of graphic processing capability make the generation of synthetic cell clusters a resource intensive task. 
When Cell cluster synthesis is compared to single-cell synthesis using equivalent constrained features and tail entities, the transformer based cluster generation performance decreases in the quantitative FID-Score from the original cell cluster patches and synthetically generated cell clusters. 

\begin{figure}[]
  \centering
  
  \subfigure{\includegraphics[height=4.45cm]{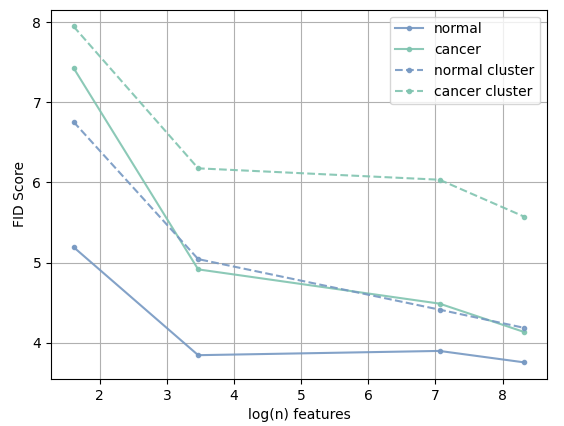}}\quad
  \subfigure{\includegraphics[height=4.45cm]{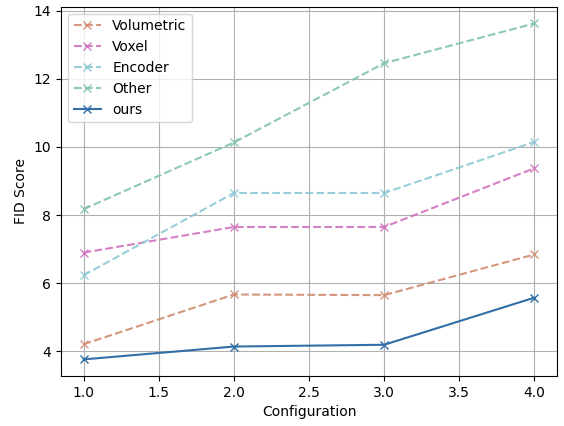}}\quad

  \caption{(a) FID Score by log(n) with n features that depicts a decreasing FID Score for a higher number of features. In (b) we compare our approach to most recent algorithms in three-dimensional synthesis such as volumetric convolution etc.}
  \label{CellClusterExamples}
\end{figure}

\section{Conclusions and future work}
We propose a novel method to synthetically generate three-dimensional cell models that allow for synthetic generation based on mutation states and the generation of entire cell clusters which enhances the understanding of cell mutation and indicates the degree and progression of cancer in a clinical assessment. The proposed algorithms achieved high quantitative and qualitative synthesis based on low Frechet-Inception scores together with visual evaluation. It enables the development of further segmentation algorithms in three-dimensional cell environments that have many advantages to reconstructed three-dimensional models such as conservation of spatial coherence and smooth shape continuity, the lack of reconstruction specific errors, and the ability to create unlimited amounts of data. Further work may focus on the synthesis of other types of tissue and cells surrounding the tissue. 

\nocite{add1, add2, add3, add4, add5, add6, add7}
%\newpage
\bibliographystyle{splncs04}
\bibliography{refs}

\end{document}